\documentclass[]{aastex631}
\usepackage{amsfonts, amsmath, amssymb}

\newcommand{\target}{CPD\,$-$29 2176}

\begin{document}

\title{Optical Properties and Variability of the Be X-ray binary \target}

\author[0000-0002-3244-682X]{Clarissa M. Pavao}
\affiliation{Department of Physics and Astronomy, Embry-Riddle Aeronautical University, 3700 Willow Creek Rd., Prescott, AZ 86301, USA}

\author[0000-0002-2806-9339]{Noel D. Richardson}
\affiliation{Department of Physics and Astronomy, Embry-Riddle Aeronautical University, 3700 Willow Creek Rd., Prescott, AZ 86301, USA}

\author[0000-0002-2919-6786]{Jonathan Labadie-Bartz}
\affiliation{LESIA, Paris Observatory, PSL University, CNRS, Sorbonne University, Universit\'e Paris Cit\'e, 5 place Jules Janssen, 92195 Meudon, France}

\author[0000-0002-1355-5860]{Herbert Pablo}
\affiliation{American Association of Variable Star Observers, 185 Alewife Brook Pkwy, Cambridge, MA, USA}

\author[0000-0002-1115-6559]{Andr\'e-Nicolas Chen\'e}
\affiliation{NSF's NOIRLab, 670 N. A'ohoku Place, Hilo, HI 96720, USA}

\begin{abstract}

Be X-ray binaries (Be XRBs) are high-mass X-ray binaries, with a neutron star or black hole orbiting and accreting material from a non-supergiant B-star that is rotating at a near critical rate. These objects are prime targets to understand past binary interactions as the neutron star or black hole progenitor likely experienced Roche lobe overflow to spin up the Be star we observe now. The stellar variability can then allow us to explore the stellar structure of these objects. 
It was recently demonstrated that the high-mass X-ray binary \target\ descended from an ultra-stripped supernova and is a prime target to evolve into an eventual binary neutron star and kilonova. We present the photometric variability from both \textit{TESS} and ASAS-SN along with the spectral properties and disk variability of the system in this paper. All of the optical lines are contaminated with disk emission except for the \ion{He}{2} $\lambda$4686 absorption line. The disk variability time-scales are not the same as the orbital time scale, but could be related to the X-ray outbursts that have been recorded by \textit{Swift}. We end our study with a discussion comparing \target\ to classical Be stars and other Be X-ray binaries, finding the stellar rotation to be near a frequency of 1.5 cycles d$^{-1}$, and exhibiting incoherent variability in three frequency groups.

\end{abstract}


\section{Introduction} \label{sec:intro}

X-ray binaries are luminous X-ray sources in the sky and consist of a compact object such as a white dwarf, neutron star, or a black hole, orbiting a normal companion star which is often evolved \citep{Biswajit2011}. There are both low-mass and high-mass X-ray binaries, and in this paper we focus on the system \target, a high-mass X-ray binary (HMXB). HMXBs consist of a massive and/or early type B or O type star companion \citep{Reig2011}. HMXBs with a Be star companion, which is a rapidly-rotating, non-supergiant B star that shows spectral emission lines in at least the Balmer lines (Be stars), and a neutron star are called Be X-ray binaries \citep[Be XRBs; ][]{2019IAUS..346..105R}. Many of the widest systems among X-ray binaries are Be systems.

Be XRBs are usually considered close systems, meaning that mass transfer from the Be star to the compact object is ongoing \citep{Reig2011}. However, since Be stars are on or near the main sequence, and the periastron distance is still large compared to the stellar radius, the Be star is not filling its Roche lobe and the mass transfer is mediated by the Be disk which is considerably larger than the star itself. In Be XRBs, the orbit typically has a moderate eccentricity \citep[e.g.,][]{Biswajit2011}. Most of the time during the Be XRBs orbit, the neutron star will be far away from the circumstellar disc that surrounds the Be star. The disc is the main source of variability in Be XRBs because it evolves on a much faster timescale than other components in the binary. The disc emits optical and infrared light, which, in turn, contaminates the magnitudes, colors and the spectral lines of the underlying star. This makes it difficult to determine fundamental stellar parameters \citet{P.Reig2016}. However, the disc can be truncated by the compact companion, as evidenced with correlations between the maximum observed intensity of the H$\alpha$ emission compared to either the orbital period \citep{2007MNRAS.377..867R,2009ApJ...707.1080A} or the semi-major axis value.

Mass transfer from the disk onto the compact object is the source of X-rays observed in Be XRBs \citep{P.Reig2016}. This equatorial disc is formed by matter expelled from the rapidly rotating Be star, although the exact mechanisms responsible for the mass ejection are still not fully understood, but may be linked to pulsation \citep[e.g., ][]{1998ASPC..135..343R, 2018A&A...610A..70B, 2021MNRAS.508.2002R}. In a Be XRB with an eccentric orbit, the neutron star can pass close to or sometimes through the disc near periastron, which causes a major disruption to the system as kinetic energy from the in-falling matter is converted to X-ray radiation \citep{Reig2011}, which can trigger variations in the Be disk. 
Be stars often have rotational velocities at or near break-up or critical velocity. Studies have shown that a sizeable amount of Be stars rotate at 70-80\% of the critical velocity \citep{Biswajit2011}. Many Be stars have been shown to have stripped companions such as sdO stars, providing evidence that they were spun up through past mass transfer in interacting binaries \citep[e.g.,][, and references therein]{2018ApJ...853..156W, 2021AJ....161..248W}.

\target\ is a fairly recently identified Be XRB. It was previously classified as B0Ve by \citet{1966AJ.....71..999F}, which was confirmed by both \citet{1983MNRAS.205..241R} and \citet{1993ApJS...89..293V}. \citet{noel} found that the binary system has a history of evolution that indicates the X-ray component was formed during an ultra-stripped supernova, which produces minimal ejecta. Additionally, the binary's orbit is similar to only one of 14 known Be X-ray binaries with published orbits in terms of both period and eccentricity.

An intriguing aspect of \target\ is that it may be associated with a magnetar-like outburst in 2016. In March 2016, \textit{Swift} BAT detected a short magnetar-like burst from an apparent new Soft Gamma Repeater, SGR 0755-2933 \citep{2016ATel.8831....1B}. Such bursts are believed to be powered by neutron star crust fractures associated with magnetic field re-arrangements in magnetars \citep{1995MNRAS.275..255T}. This object was monitored by {\it Swift} \citep[e.g.,][]{2016ATel.8868....1A} which placed an upper limit on pulsed fraction of the soft X-ray flux at 15\% when it returned to a quiescent state. Furthermore, no pulsations were found at radio wavelengths \citep{2016ATel.8943....1S}. While the object seems to possess many properties of magnetars, as evidenced by these early reported observations, \citet{2021A&A...647A.165D} question the classification of the source as a magnetar and suggest it is a Be XRB with a normal neutron star based on an optically bright counterpart. 
The probability of a chance alignment in the sky of a massive star and a magnetar or isolated neutron star is low \citep{noel}, especially given the scarcity of these compact objects, making \target\ a Be XRB given the orbit presented by \citet{noel}.
It is thus possible that the companion to the Be star is a magnetar rather than a normal neutron star. While Younes et al.~(in prep.) find the source to be typical for a magnetar, we also caution that the astrometry of the source was accomplished with \textit{Chandra} data, while the magnetar-like outburst was only observed with \textit{Swift}. Instumental differences could still mean that the neutron star present in \target\ is a standard neutron star and unrelated to the magnetar-like outburst.
neutron star present is not a bona fide magnetar. 
\citet{2023arXiv230713308N} showed the long-term X-ray light curve of \target\ which shows the first, strong outburst, a potential rise towards a second outburst 300-400 d later, and two additional outbursts separated by $\sim 300-400$ d after a longer observational gap. 

In this paper, we analyze the series of optical spectra presented for the orbital motion in \citet{noel} along with archival photometric time-series that have been recorded. We describe the observations in Section 2, and describe the photometric variations in Section 3. Section 4 describes the optical spectrum in depth with the variability in the disk emission lines shown in Section 5. We discuss our findings in Section 6 and conclude our study in Section 7. 

\section{Observations}

\subsection{Photometry}
The All Sky Automated Survey (ASAS) and the All Sky Automated Survey for Supernovae (ASAS-SN) observed \target\ over a large period of time. The ASAS data was taken from November 2000 to December 2009 with the $V-$band filter \citep{1997AcA....47..467P, 2003AcA....53..341P}. We examined the ASAS-SN with the $V$-band filter that were taken from March 2016 to April 2018 \citep{2014ApJ...788...48S, 2017PASP..129j4502K}, but found that these data were saturated and thus unreliable. 

The \textit{Transiting Exoplanet Survey Satellite} (\textit{TESS}) observed \target\ in four $\sim$27 d sectors of its mission thus far: Sectors 7, 8, 34, and 61. We extracted the flux from the full-frame images for these observations. These sectors correspond to times around January 2019, February 2019, January 2021, and February 2023 respectively. The \textit{TESS} photometric points have a precision of a few parts per 10,000 and are taken with a cadence of once every 30 minutes (sectors 7 and 8), 10 minutes (sector 34), and 200 seconds (sector 61).

\begin{figure}
    \centering
    \includegraphics[width=5.5in]{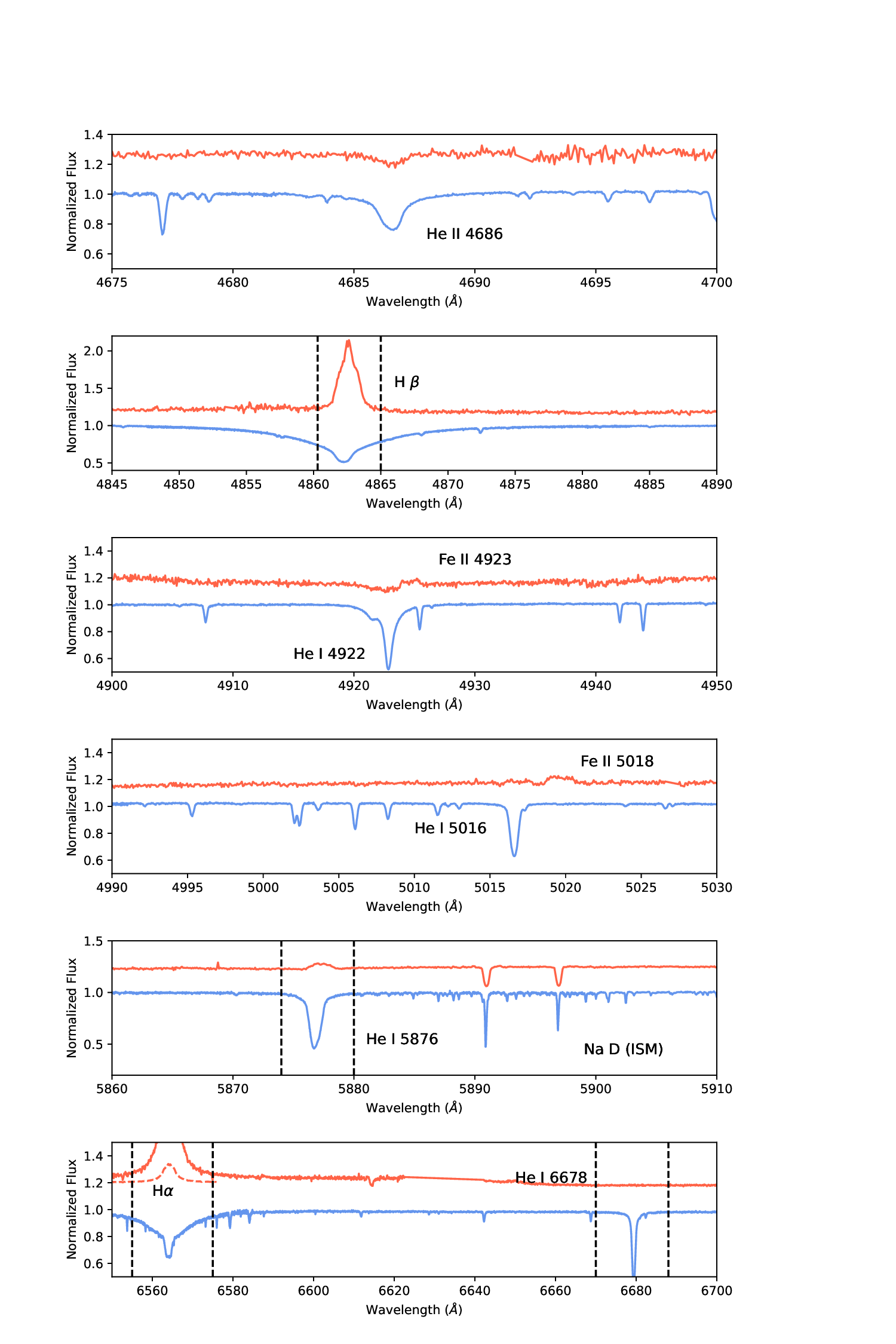}
    \caption{The optical spectrum of \target\, shown in red (upper spectrum), with a comparison to the B0V star $\tau$ Sco from the ESPaDOnS archive shown as a blue line. The spectra are offset for clarity, and the spectrum of $\tau$ Sco was shifted in radial velocity to match the $\gamma$-velocity of the binary presented by \citet{noel}. In the lower panel, we also show the H$\alpha$ profile scaled down by a factor of 10 for clarity.}
    \label{fig:spec_atlas}
\end{figure}

\subsection{Optical Spectroscopy}

We collected optical spectroscopy of \target\ with the CTIO 1.5 m telescope and echelle spectograph CHIRON from 2016-2021. \target\ was observed 16 times between 2018 September 30 to 2019 May 20 (typically 3 spectra per night). It was observed again 2020 December 13 to 2021 January 25 (typically 2 spectra per night). The spectrograph has a spectral resolution of R $\sim$ 28,000 with a spectral range of 450-870 nm fixed. The spectra observed for each night were co-added and have a S/N of $\sim$50 in the continuum near the \ion{He}{1} 5876 \AA\ line. All spectra were normalized around the unit continuum. We analyzed the \ion{He}{1} $\lambda$5876 \AA\ line, \ion{He}{1} $\lambda$ 6678 \AA\ line, H$\beta$ line, and H$\alpha$ line in the disk variability analysis presented here. These observations were also analyzed by \citet{noel}, but only with the (photospheric) \ion{He}{2} $\lambda$4686 \AA\ line.

\section{The optical spectrum of \target}

\citet{noel} presented the discovery of the orbital motion of \target\ using optical spectroscopy obtained with the CHIRON spectrograph. Using these observations, we created an average spectrum of these spectra in order to increase the signal-to-noise ratio as well as to identify any lines of interest in the optical spectrum. As the semi-amplitude of the orbital motion is only a few km s$^{-1}$, we did not shift the individual observations to the same $\gamma$-velocity as the lines are wider than the orbital motion amplitude. The resulting average spectrum is shown in Fig.~\ref{fig:spec_atlas}, along with a comparison to a B0V standard star, $\tau$ Sco \citep{1973ARA&A..11...29M}, taken with the ESPaDOnS spectrograph and obtained from the PolarBase archive \citep{1997MNRAS.291..658D, 2014PASP..126..469P}. We use this star as \target\ has been classified as a B0V star several times in the literature \citep{1966AJ.....71..999F, 1983MNRAS.205..241R, 1993ApJS...89..293V}.

In Fig.~\ref{fig:spec_atlas}, we note that the two strongest lines in the spectrum of \target\ are the
Balmer emission lines of H$\alpha$ and H$\beta$. These lines show a classic ``wine bottle" shape with a narrow top on top of a broader emission. According to the models of Be stars, and the geometries presented by \citet{1996A&AS..116..309H} and \citet{Riv2013}, this corresponds to a near-pole on geometry for the Be disk. This agrees with the small semi-amplitude of the orbit and the proposed scenario presented by \citet{noel}. 

The optical \ion{He}{1} lines appear nearly absent from the spectrum of the star, as seen around the positions of 4921, 5016, 5876, and 6678 \AA\ in Fig.~\ref{fig:spec_atlas}. Given the emission profile that is evident for the 5876 line and weak double-peaked emission profiles present at the other \ion{He}{1} line positions, we suggest that these lines are present in the spectrum of the star, but have emission that fills the absorption for these lines. Near the \ion{He}{1} 5876 \AA\ emission line, we also note the two strong \ion{Na}{1} $D$ absorption lines that are interstellar in nature.

The disk is also present and somewhat stronger in its emission for some \ion{Fe}{2} lines in the spectrum. Evident in these observations shown in Fig.~\ref{fig:spec_atlas} are the \ion{Fe}{2} lines that blend with \ion{He}{1} lines, namely at 4924 and 5018 \AA. These lines also show a weak but broad double-peaked geometry that is of similar width to the \ion{He}{1} 5876 \AA\ emission line. The presence of \ion{Fe}{2} is only in emission, in contrast to ions such as \ion{He}{1} which is seen in absorption for a standard B0V star and thus the observed \ion{He}{1} lines are a mixture of both photospheric absorption and disk emission. 
\ion{Fe}{2} emission can only arise from relatively cool circumstellar gas. This condition can be met in disks around early Be stars if the inner disk is sufficiently dense that it shields disk material from the ionizing stellar UV radiation. The result is a relatively cool pocket of gas at a few stellar radii \citep[e.g., Fig. 9 in ][]{2021ApJ...912...76M}, which is consistent with the wide doppler broadening of the \ion{Fe}{2} emission observed in \target.

Lastly, we also show the weak \ion{He}{2} absorption line at 4686 \AA. This is the strongest \ion{He}{2} line in the optical range and represents the hydrogen-like $4\rightarrow 3$ transition. 
The spectrum of \target\ is dominated by strong emission lines formed in the Be decretion disk. The top panel of Fig.~\ref{fig:spec_atlas} shows the region surrounding the \ion{He}{2} $\lambda$4686 line. \target\ shows a weak absorption line as was discussed by \citet{noel}. There are multiple other weak, narrow absorption lines seen in the spectrum of the B0V standard star $\tau$ Sco. Likely these lines are either broadened by rotation of the Be star in \target\ or filled by emission such that they are not present in our observations. 

\section{Photometric Variability}

\begin{figure}
    \centering
    \includegraphics[width=5.5in]{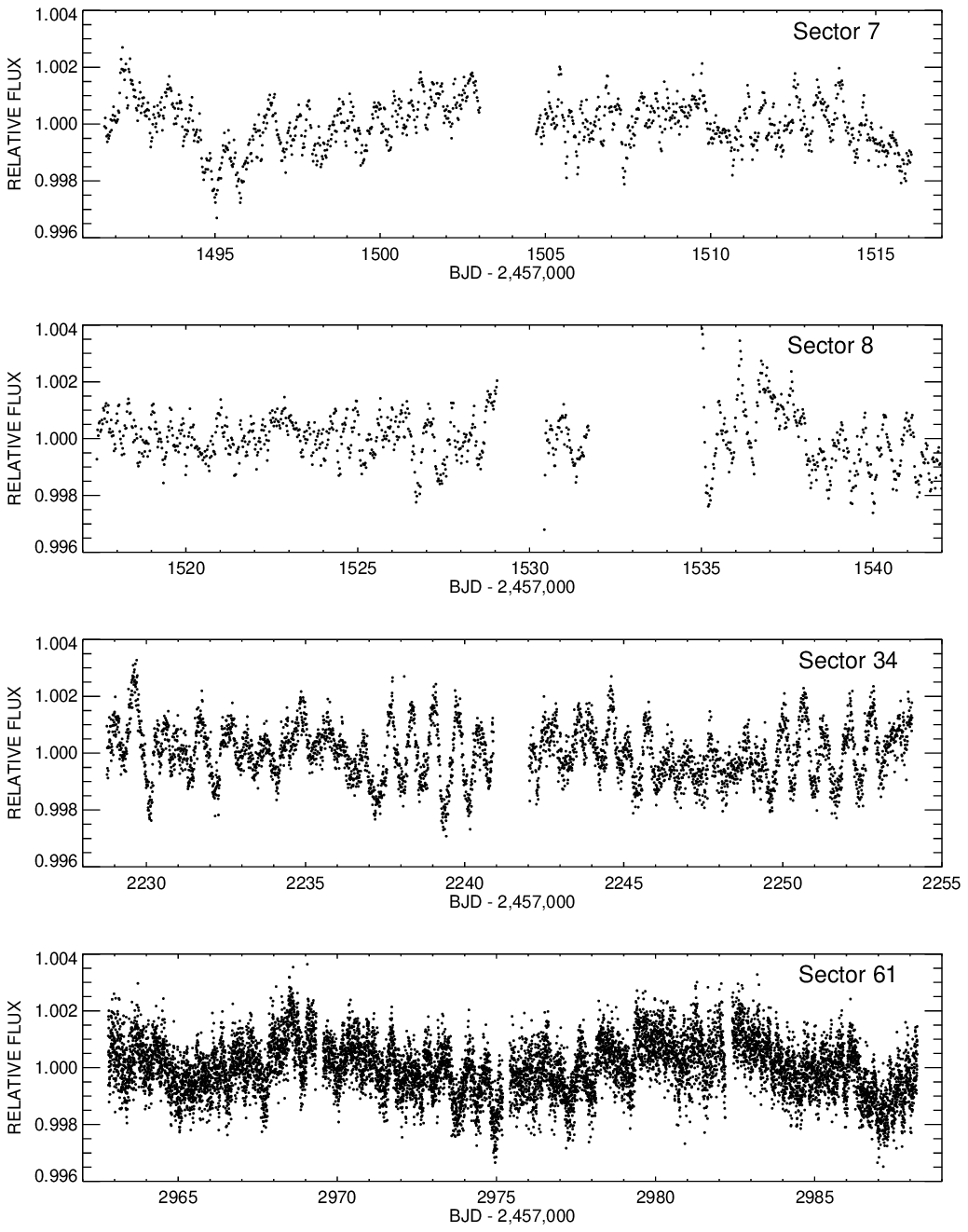}
    \caption{The short-term variability of \target\ as observed by \textit{TESS}. Each of the four sectors of \textit{TESS} photometry is shown in a separate panel as indicated.}
    \label{fig:tessLC}
\end{figure}

The \textit{TESS} observations were collected over four sectors of the \textit{TESS} mission. Each year with observations was collected with a higher cadence time-series than the previous, as is evident in the light curve presented in Fig.~\ref{fig:tessLC}. The light curve variation is at the $<$1$\%$ level, and is primary stochastic but with some structure.

\begin{figure}
    \centering
    \includegraphics[angle=90, width=\columnwidth]{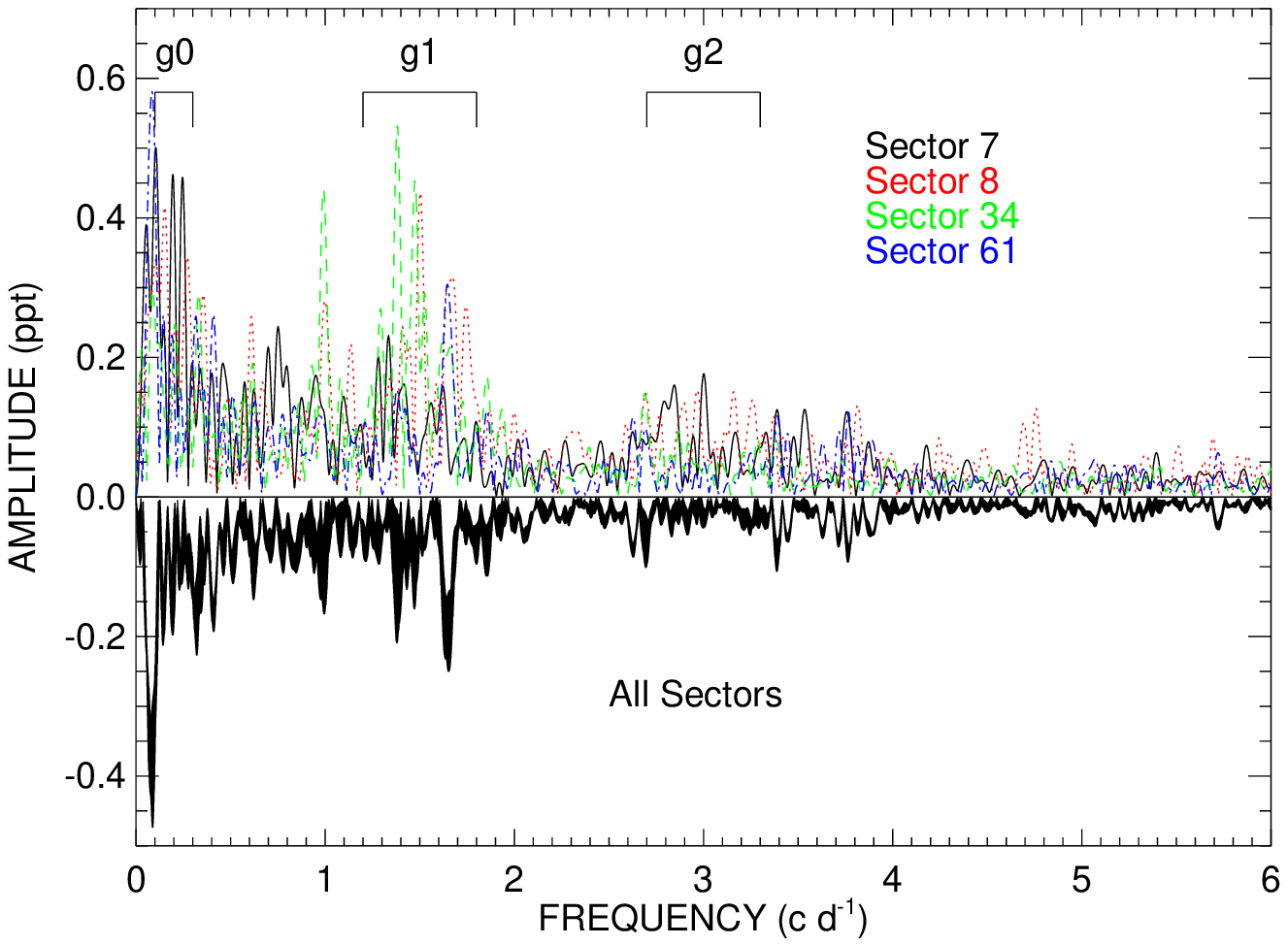}
    \caption{The Fourier analysis of each sector of the \textit{TESS} data shown in different colors and line styles. The negative-valued black spectrum represents the Fourier analysis of the light curve containing all four sectors. We also note some frequency groups with the designations g0, g1, and g2 on the top.}
        \label{fig:tessFT}
\end{figure}

In Fig.~\ref{fig:tessFT}, we present a Fourier analysis of the \textit{TESS} light curve calculated with {\tt period04} \citep{2005CoAst.146...53L}, with each sectors analyzed independently. We estimate the typical noise level for these periodograms to be about 0.014 parts per thousand. Our noise estimate is a typical value for the Fourier amplitude spectrum at high frequencies ($f > 6$ cycles d$^{-1}$) where no coherent variations were present in the Fourier spectrum.

Of note in these Fourier periodograms are three regions that show power excess, representing three frequency groups in the light curve analysis. The first, which we label `g0' in Fig.~\ref{fig:tessFT}, is a low frequency group of peaks near 0.1 cycles d$^{-1}$. This slow variability is very likely real and not just badly detrended systematics from the satellite but it is difficult to exactly pin down their nature. For example, these could represent difference frequencies between various oscillations in the light curve driven by pulsations, or they could represent slow, long-term evolution of the disk surrounding the Be star that is driven by a variable mass-loss rate into the disk. Such frequency groups are common in Be stars \citep[e.g., ][]{2022AJ....163..226L, 2018A&A...613A..70S, 2016A&A...588A..56B}.

We also label a strong group of frequencies `g1' in Fig.~\ref{fig:tessFT}, which is centered near 1.5 cycles d$^{-1}$. The strongest peak in this region varies from sector to sector, and the Fourier analysis of the entire \textit{TESS} light curve seems to show a frequency near the high-end of this group near 1.7 cycles d$^{-1}$, but note that the frequency group spans from $\sim 1.3$ to $\sim 1.8$ cycles d$^{-1}$. We would need a longer time-series to resolve if there are any genuine stable, coherent frequencies in the regime. However, the amplitude of these frequencies is very low, near about 0.2--0.4 parts per thousand, likely below the threshold for ground-based data sets. Similarly, there appears to be a lower-amplitude group just above the level of significance near 3 cycles d$^{-1}$ group at `g2'. This frequency group pattern, with `g2' located at $\sim$ twice the frequency of `g1' is seen in the large majority of Be stars \citep{2022AJ....163..226L, 2018A&A...613A..70S}. In the Fourier analysis of all four sectors of \textit{TESS} data, we see the strongest signal in g2 is at $\sim$3.4 cycles d$^{-1}$, where we also see a similar peak in each of the four sectors when analyzed separately. This frequency is about 2$\times$ the strongest frequency in g1 seen in the Fourier analysis of all four sectors.

\begin{figure}
    \centering
    \includegraphics[angle=0]{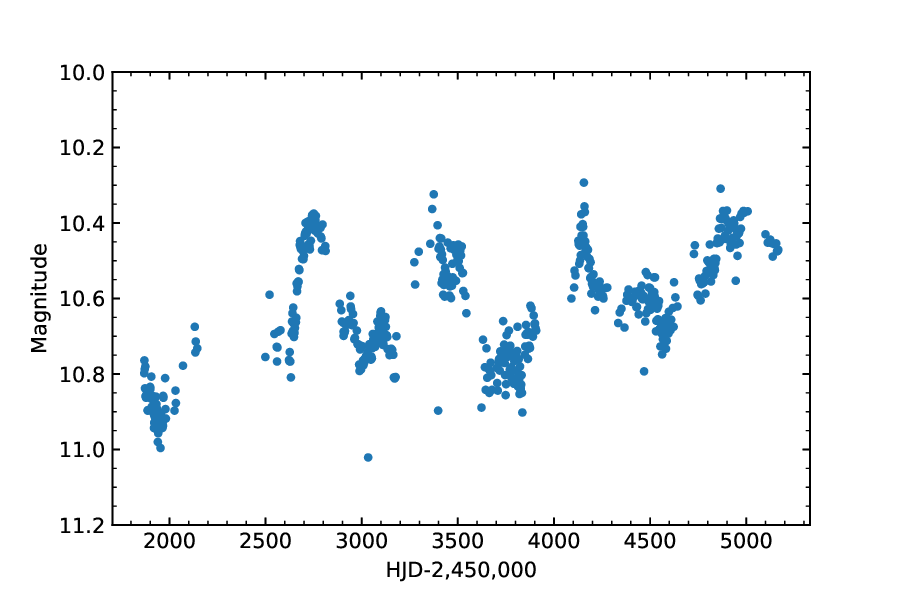}
    \includegraphics[angle=0]{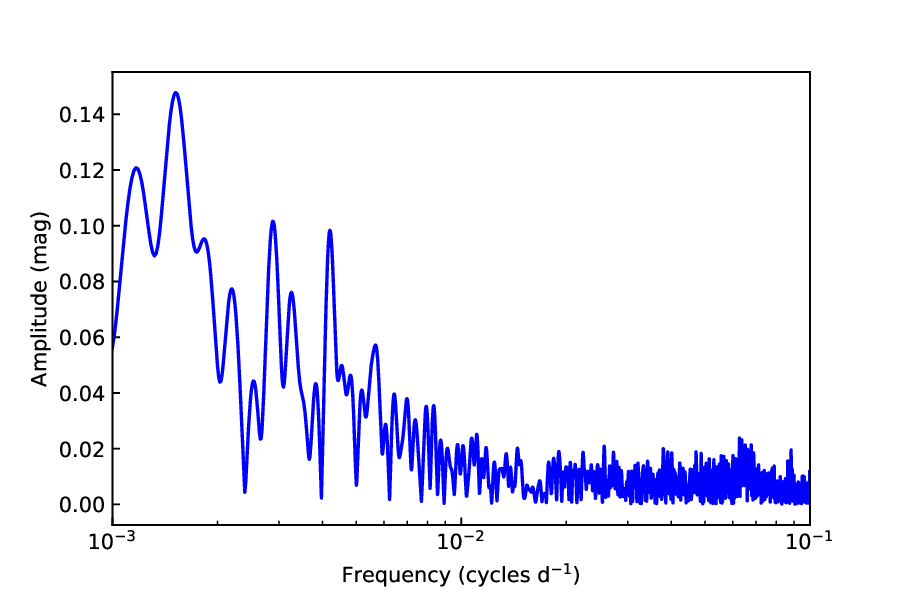} 
    \caption{The long-term photometric variability of \target\ as observed with ASAS as well as the Fourier spectrum of the light curve. The Fourier spectrum shows a time-scale of 600--700 d, but the light curve is not periodic as it does not phase onto these periods.  }
    \label{fig:Photometry}
\end{figure}

The long-term variabilty of \target\ is best explored with the time-series from ground-based surveys such as ASAS, which we show in Fig.~\ref{fig:Photometry}. The photometric data do not show evidences of eclipses but do show long term variations similar to other Be stars \citep{2017AJ....153..252L,2018MNRAS.479.2909B} including Be XRBs \citep{2015A&A...574A..33R,2023A&A...671A..48R}.
In the ASAS data, we see multiple peaks in the light curve with a timescale of $\sim600$ d. We examined a Fourier spectrum of the ASAS photometry using {\tt period04} \citep{2005CoAst.146...53L} and find a signal with a time-scale of 600-800 d, but this is not periodic as it does not phase well. We also subtracted long-term trends and found no evidence of the orbital period presented by \citet{noel}. The Fourier spectrum is also shown in Fig.~\ref{fig:Photometry}.

\section{Long-term Variability of Disk lines}

\begin{figure}
    \centering
    \includegraphics[angle=0,width=0.7\columnwidth]{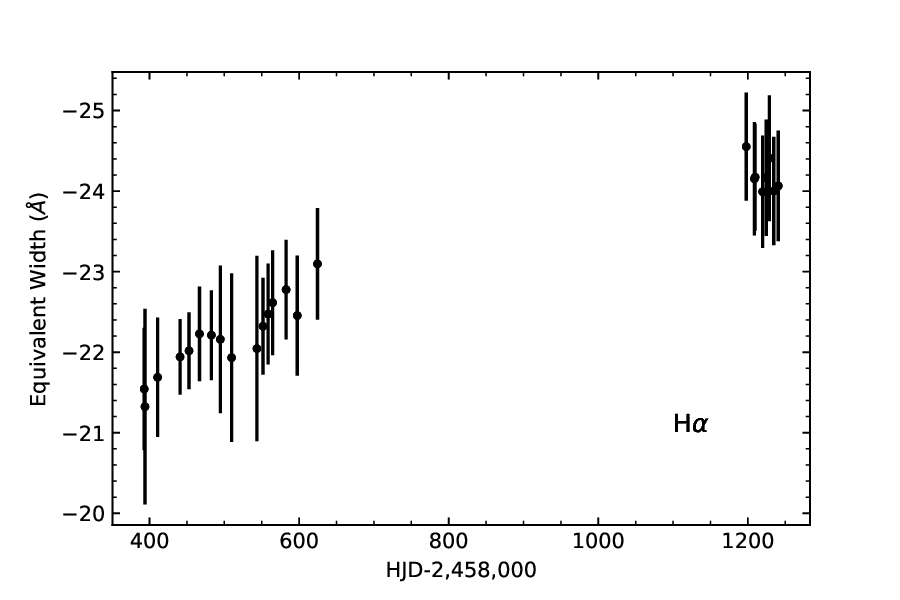}
    \includegraphics[angle=0,width=0.7\columnwidth]{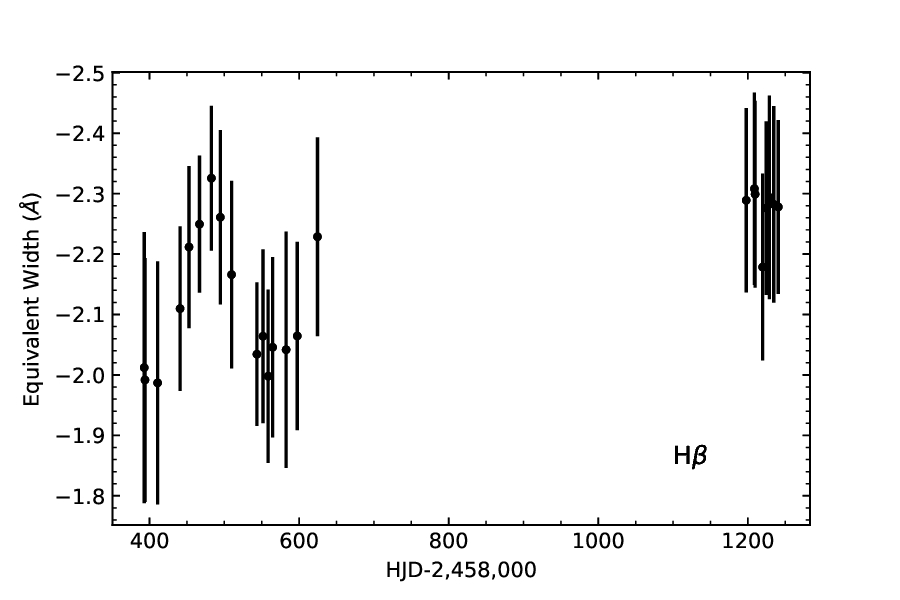}
    \caption{Equivalent Width measurements of \target\ for the optical hydrogen lines.}
    \label{fig:equivalent width - hydrogen}
\end{figure}

We measured the equivalent width of several disk emission lines in the CHIRON spectroscopy. We start this discussion with the H$\alpha$ and H$\beta$ lines, which we show in Fig.~\ref{fig:equivalent width - hydrogen}. The H$\alpha$ line is a spectral line in the Balmer series that is most influenced by the circumstellar environment \citep[e.g.,][]{B}. To measure the H$\alpha$ line, we normalized our spectra by the continuum before integrating the flux 
between 6552-6580 \AA\ ($-500$ to $+800$ km s$^{-1}$ to 
secure a continuum measurement on both sides of the profile).
From the equivalent widths of H$\alpha$, we observe a weak, but steady increase of the emission strength. The H$\beta$ line was measured by integrating between 4859-4865 {\AA} ($-130$ to $+250$ km s$^{-1}$). These ranges were picked to reach the continuum as well as to avoid using multiple echelle orders in the measurement. 
The H$\beta$ equivalent width plot, Fig. 5, shows a small increase
in the early observations that is not present in the H$\alpha$ observations. For both data sets, the slow increases in equivalent width are longer than the orbital period of 59.6 days by \citet{noel}. The equivalent widths of both H$\alpha$ and H$\beta$ can be found in Table \ref{measurements} along with the exposure times and the signal-to-noise ratio of the continuum near the echelle spectrograph's blaze maximum.

\begin{figure}
    \centering
    \includegraphics[angle=0,width=0.7\columnwidth]{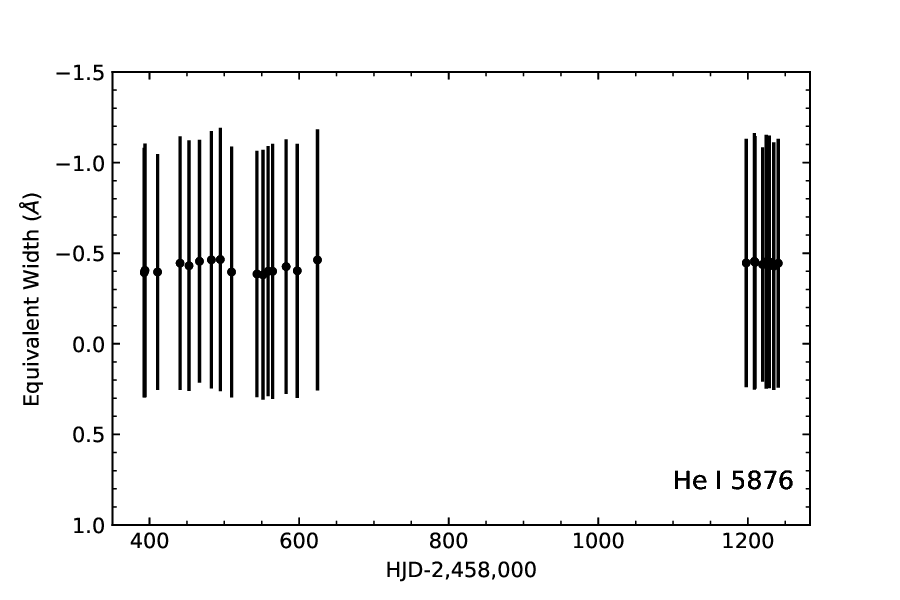}
    \includegraphics[angle=0,width=0.7\columnwidth]{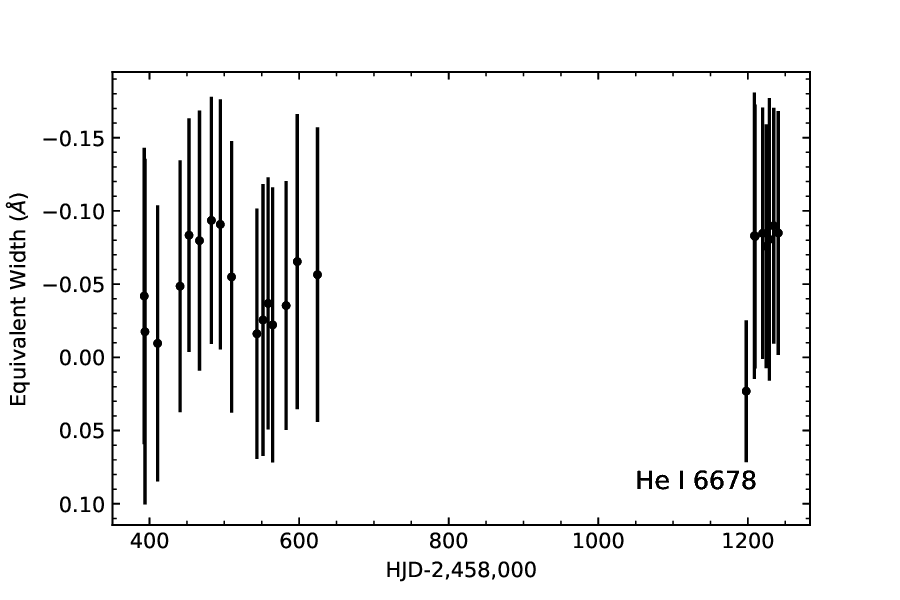}
    \caption{Equivalent Width measurements of \target\ for the two strong optical helium lines we measured. }
    \label{fig:equivalent width - helium}
\end{figure}

We also measured the equivalent width of two prominent helium lines in the spectra of \target, namely  \ion{He}{1} ${\lambda}$5876 {\AA} and  \ion{He}{1} ${\lambda}$6678 {\AA} which is shown in Fig.~\ref{fig:equivalent width - helium}. These two helium lines are prominent in hot Be stars and tend to be in emission when the inner disk is dense. Once the spectrum was normalized, we measured the equivalent widths of the profiles over the range from 5874-5880 {\AA} and over 6670-6688 {\AA} for the 5876 and 6678 \AA\ lines respectively. Neither of these lines show any statistically significant trends in their variations.

\section{Discussion}

The \textit{TESS} photometry, ASAS photometry, and ground-based optical spectroscopy have provided us a first look into the variability of \target. This is an important system given the recent findings of \citet{noel}, who showed that the \ion{He}{2} 4686\AA\ absorption line shows evidence for a circular orbit with a period of $\sim$59\,d with a mass function that supports a neutron star in a nearly face-on orbit. 

Thus far, we have shown that the optical light curve as observed with ASAS exhibits quasi-regular outbursts with a time-scale of $\sim$600 d. The optical spectroscopy, while too sparse to examine such long time scales, did not show signs of large disk variability with this same timescale, but did show some trends in the hydrogen line profiles. Neither these spectroscopic variations nor the photometric variations are coherent on the orbital time scale of $\sim$60 d.

\target\ has several properties that are comparable to classical Be stars. Its H$\alpha$ and H$\beta$ emission line profiles, shown in Fig.~\ref{fig:spec_atlas}, have a ``wine bottle" profile, which is consistent with a pole-on geometry of the disk. The binary evolution scenario presented by \citet{noel} would imply that the orbit and the Be star geometry would need to be co-planar. Thus, the disk emission line profiles support the evolutionary picture and allows us to consider the current evolutionary state of the system in comparison to other Be X-ray binaries. The disk profiles for \target\ are not extremely variable in profile morphology or strength as evidenced in our analysis. As such, \target\ with its circular orbit may represent a Be XRB that could be used to study a more stable geometry of a Be disk in a Be XRB. We also note that there could be an observational bias for \target\, as the face-on geometry could also make this system appear more stable than other systems seen edge-on, at least in the azimuthal distribution of matter in the disk. In contrast, the ASAS data show regular variations that indicate mass ejections into the disk. Such photometric changes without clear spectroscopic changes show us how face-on Be stars can blind us to azimuthal changes in density perturbations while we are more sensitive to observing these mass ejections.

The majority of Be stars seem to pulsate in non-radial low order sectoral (where $l = |m|$) g modes where gravity is the restoring force \citep{Rivinius2003}, similar to the class of Slowly Pulsating B (SPB) stars \citep{DeCat2002}. The photometric amplitude of such modes depends on the stellar inclination angle. At low inclinations, sectoral modes have low amplitudes due to geometric cancellation over the integrated stellar surface, to the point where they will not impart any photometric signal at $i = 0^{\circ}$. In \target, there are not any obvious coherent peaks that stand out in the frequency spectrum. This is not unusual for early-type Be stars \citep[e.g.][]{2020MNRAS.498.3171N, 2022AJ....163..226L}. Multiple factors may contribute to the lack of obvious coherent oscillation, including the low inclination angle, lower intrinsic g mode amplitudes in the hottest Be stars, and the large number of closely spaced signals which may mask the presence of coherent modes without a sufficiently long time baseline. 

While coherent oscillations are generally associated with stellar pulsation, Be stars in particular also often display incoherent oscillations consistent with pulsation (often, but not always, in addition to coherent modes). This is most evident during times where the Be star is actively ejecting mass \citep{2018A&A...613A..70S,2022AJ....163..226L}. While mass ejection events are commonly marked by a net increase in brightness (which are not seen in the TESS light curves for \target), this is not always the case. If the inner disk is dense and optically thick, small additions of material will not increase the net emergent flux. This situation may be realized in \target, since the spectra indicate a strong disk was present at all observed times including during TESS sectors 7, 8, and 34 (note that our spectroscopic coverage does not extend to sector 61). 

Given the low inclination angle, it is unlikely that such signals could be attributed to any circumstellar activity such inhomogeneous ejecta transiting the stellar disk, although this cannot strictly be ruled out with photometry alone. One plausible explanation for these incoherent signals are Rossby waves (r modes), which are expected and observed in rapid rotators \citep{VanReeth2016,Saio2018}. Rossby waves are necessarily retrograde and tend to form frequency groups just below the rotation frequency (for $m = 1$) and just below twice the rotation frequency (for $m = 2$), and so on with decreasing amplitude with higher $m$. During mass ejection, the rotational flow on the stellar surface would be disrupted, potentially generating r modes \citep[as conjectured in][]{Saio2018}. In this scenario, if mass ejection rates vary over time (which seems to be the case in all early-type Be stars), presumably any subsequently generated r modes would not be coherent over long periods of time (e.g., from sector to sector in TESS).

$\gamma$ Cas (B0.5\,IVe) was the first discovered Be star \citep{1866AN.....68...63S} and is the prototype of a subgroup of classical Be stars showing X-ray emission \citep{2020MNRAS.493.2511N}. However, the X-rays of $\gamma$ Cas stars are distinguished by being hard, thermal, and of intermediate strength between normal OB stars and Be XRBs \citep{2006A&A...454..265L}, and the X-ray mechanism is still unknown. The optical properties of $\gamma$ Cas and \target\ are similar, with the main differences being due to the binary separation and the inclination angle. In both systems the binary orbits are consistent with being circular, $\gamma$ Cas having P$_{\rm orb}$ = 203.5 d \citep{2012A&A...540A..53S, Nermravov}, and \target\ having P$_{\rm orb}$ = 59 d \citep{noel}. The disk of $\gamma$ Cas has an inclination of $(41\pm4)^\circ$ based on interferometric measurements \citep{2012A&A...545A..59S} while \target\ has an inclination between 10$^\circ$ and 15$^\circ$ \citep{noel}. From the ASAS photometry (Fig.~\ref{fig:Photometry}), the $max - min$ variation of \target\ in $V_{mag} \approx 0.56$, which is among the most extreme values for classical Be stars \citep{2018MNRAS.479.2909B}. The geometry of \target\ is favorable for high amplitude flux variations since face-on disks have a maximum projected area, but even so the disk of \target\ must have a high density at times of maximum brightness \citep{2012ApJ...756..156H}. This, perhaps, can be aided by the binary companion whose tidal forces cause disk material to accumulate rather than flow outwards, as the $V$ band flux only probes the more inner regions of the disk \citep{2015MNRAS.454.2107V}. The magnitude of the $V$ band flux variation in $\gamma$ Cas is similar, but on much longer timescales (decades) compared to \target\, most likely due to mass loss in $\gamma$ Cas being relatively continuous while in \target\, at least during the ASAS observations, mass ejection was episodic.

\target\ also shows other similarities to $\gamma$ Cas. The frequency groups derived from the \textit{TESS} photometry (Fig.~\ref{fig:tessFT}) are similar to those seen in $\gamma$ Cas \citep{Labadie}. In contrast though, $\gamma$ Cas does seem to show one coherent frequency in its equivalent of the `g2' group during the TESS epochs, but even so this frequency was not detected in earlier observations \citep{HenrySmith} which instead found a, at the time, coherent signal at 1.2158 d$^{-1}$ which was not apparent in the TESS data. In Be stars with groups like this, the stellar rotation frequency should be approximately the difference between the two major frequency groups \citep{2018A&A...613A..70S}, so the rotation frequency for \target\ should be near 1.4--1.6 cycles d$^{-1}$.
This qualitatively makes sense if you imagine the frequencies in the `g1' group are a combination of prograde $g-$modes and retrograde $r/g-$modes, which will then bracket the rotational frequency $f_{\rm rot}$, although we are not certain what the \textit{TESS} frequencies represent yet, but this could become clearer with future coordinated spectroscopic observations at the time of \textit{TESS} photometric observations or other space photometry.

\begin{figure}
    \centering
    \includegraphics{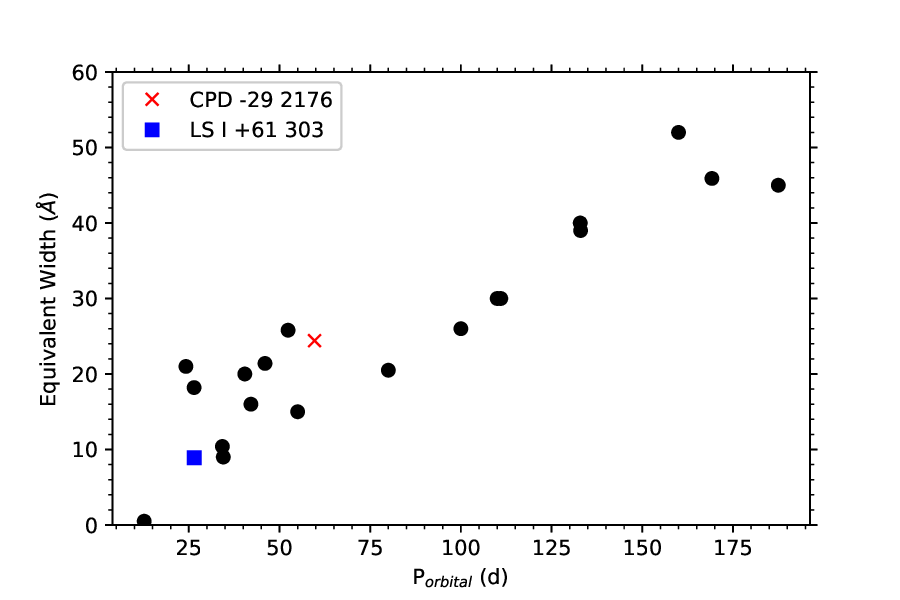}
    \caption{The maximum observed H$\alpha$ equivalent width compared to the orbital period for Be XRBs. We highlight \target\ and LS I +61 303.}
    \label{fig:porb-ew}
\end{figure}

The orbital period of \target\ and our observed H$\alpha$ strength fits into the relationship between orbital period and the maximum observed H$\alpha$ equivalent width as originally shown in \citet{1997A&A...322..193R}, and subsequently in \citet{2007MNRAS.377..867R}, \citet{Reig2011}, \citet{2009ApJ...707.1080A}, \citet{2015MNRAS.452..969C}, and \citet{2016A&A...590A.122R}. We show the placement of \target\ in the population of Be XRBs in Fig.~\ref{fig:porb-ew}.
The relationship between the H$\alpha$ equivalent width and orbital period has been modeled by theory developed by \citet{2002MNRAS.337..967O}. 
The primary reason for this trend is that tidal forces from the companion truncate the Be disk, inhibiting radial growth and resulting in a shallower radial density gradient. The smaller geometric disk area then results in a reduced H$\alpha$ equivalent width. In eccentric binary orbits, the observed emission levels are often phase dependent, but in the circular case a $\sim$steady state is realized. Some disk material accretes onto the companion, while some may be ejected from the system.
Similar modeling and observational results have been shown for hot, stripped companions for classical Be stars \citep[e.g.,][]{2017ApJ...836..112D, 2019ApJ...885..147K}.

In Be stars with $\sim$stable disks and a binary companion in a circular (prograde) orbit, spiral density waves are predicted \citep{2016MNRAS.461.2616P, 2018MNRAS.473.3039P}.  Although referred to as an $m=2$ wave in the disk, the leading arm of the pattern (nearer to the companion) is stronger, introducing an asymmetry \citep{2020MNRAS.497.3525C}. The most conspicuous observable is a phase-locked pattern of variability in prominent emission lines, e.g. H$\alpha$ \citep[e.g.]{2013A&A...560A..30Z, Chojnowski2018, 2020A&A...639A..32H}. No such pattern is seen in the H$\alpha$ line of \target\ when phased to the orbital period. However, the existence of a tidally-induced $m=2$ wave cannot be ruled out since the main observational signatures become (perhaps negligibly) small at low inclinations and the number of spectra is small for detecting such a signature. Further, the stronger overall H$\alpha$ emission at later epochs (Fig.~\ref{fig:equivalent width - hydrogen}) makes direct comparisons over the full time baseline difficult. We note that \citet{2010ApJ...724..379M} described the H$\alpha$ variability of LS\,I $+$61$^\circ$ 303 ($P_{\rm orb} = 26.4960$ d, $e=0.537\pm0.034$; also highlighted in Fig.~\ref{fig:porb-ew}). Over 35 consecutive nights of spectroscopic data, \citet{2010ApJ...724..379M} found a brief time prior to apastron that has a strong excess on the red side of the H$\alpha$ profile when the neutron star moves almost directly away from the observer. \citet{2010ApJ...724..379M} interpret this emission as evidence that an induced spiral density wave in the Be disk extends across the full binary separation at these phases and thus shows that these sorts of waves can exist in Be XRBs, although the face-on geometry of \target\ may make this harder to observe directly.

All of these observables in \target\ show a consistent picture of a nearly pole-on Be star with a compact object in a circular or low-eccentricity orbit. The pole-on nature is supported by the Be emission line morphologies, the apparent lack of coherent oscillations from \textit{TESS}, and the orbital information presented by \citet{noel}. The ground-based optical light curve shows regular disk activity during the times of the ASAS project, with outbursts occurring every 600-800 d without periodicity. In the X-ray regime, the light curve shown by \citet{2023arXiv230713308N} shows possible X-ray outbursts at time scales of $\sim 300$ d. While these time-scales differ by about a factor of two, we also note that the two data sets were not contemporaneous so the disks in the system could have been very different at these two times. The system, while sharing many properties with the Be XRB LS\,I that was also speculated to have a magnetar companion, is distinct in that the circular orbit geometry brings about more stability than seen in more eccentric systems.

\section{Conclusions}

In this paper, we presented an analysis of both the spectroscopic properties and variability, along with the photometric variability of \target. Our findings can be summarized with the following points:
\begin{itemize}
    \item \target\ has a spectrum that is consistent with a nearly pole-on B0Ve star. Such an inclination provides support for the evolutionary picture presented by \citet{noel}, who found that the previous supernova required an ``ultra-stripped" supernova and presents the system as a kilonova progenitor.
    \item The high-precision \textit{TESS} photometry shows three groups of frequencies. One is at low frequency (long time scales) and likely is related to slow, long-term variability of the Be disk. With the properties of \textit{TESS} data, this refers to a time-scale of 10--20 d, although we don't see the exact same properties with the ground-based ASAS data, but rather the ASAS strength is at longer time scales. The second frequency group near 1.5 cycles d$^{-1}$ was seen in the \textit{TESS} data and does not have a stable frequency over the four sectors of data collected thus far. This group likely represents stellar oscillations. The final frequency group is near 3 cycles d$^{-1}$ and is probably related to the second group mentioned but with a higher azimuthal order. The inferred rotational frequency from these oscillation groups is about 1.5 cycles d$^{-1}$, consistent with other Be stars. 
    \item The disk around \target\ is fairly stable across our spectroscopic observations spanning about $\sim$2 years. There may be long-term photometric evolution of the system as seen with ground-based photometric surveys (Fig.~\ref{fig:Photometry}), but these time scales are largely not yet covered with spectroscopy for confirmation. The variations in the ASAS photometry could be explained by disk outbursts causing the inner disk density to vary significantly as the amplitude is high so these were not minor events. The strength of the optical \ion{He}{1} lines is relatively constant while the H$\alpha$ profile shows some strengthening with time (Figs.~\ref{fig:equivalent width - hydrogen} and \ref{fig:equivalent width - helium}).
    \item The strength of the H$\alpha$ profile fits the trend of comparing the equivalent width of H$\alpha$ with the orbital period for Be XRBs (Fig.~\ref{fig:porb-ew}). This trend is supported by modeling of disk truncation by the compact companion for other Be XRBs.  
\end{itemize}

\target\ offers a prime laboratory to study Be XRBs with circular orbits. Such systems are seemingly rare, and we note that the system LS\,I which we compared \target\ to has both high eccentricity \citep{2009ApJ...707.1080A} and was ejected from the cluster IC 1805 \citep{2004A&A...422L..29M}, meaning that while these systems both have shown evidence of magnetars as their compact companion, the evolutionary paths to form these systems are likely very different. Future studies of \target\ will likely allow us to understand its evolution in detail and place constraints on the formation of binary neutron star systems.

The spectroscopy from CTIO was
collected through the NOIRLab program nos.\ 2018B-0137 and 2020A-0054. This research has
used data from the CTIO/SMARTS 1.5m telescope, which is operated as part of the SMARTS
Consortium by RECONS (www.recons.org) members T. Henry, H. James, W.-C. Jao and
L. Paredes. At the telescope, observations were carried out by R. Aviles and R. Hinojosa. This paper includes data collected by the TESS mission. Funding for the TESS mission is provided by the NASA's Science Mission Directorate.
We thank Pablo Reig for data in Fig.~\ref{fig:porb-ew}.

C.M.P. acknowledges support from the Embry-Riddle Aeronautical
University’s Undergraduate Research Institute and the Arizona Space Grant. This research was
partially supported through the Embry-Riddle Aeronautical University’s Faculty Innovative
Research in Science and Technology (FIRST) Program and through NASA grant 80NSSC23K1049. The work of ANC is supported by NOIRLab, which is managed by the Association of Universities for Research in Astronomy (AURA) under a cooperative agreement with the National Science Foundation.

\vspace{5mm}
\facilities{TESS, ASAS, CTIO:1.5m}

\software{astropy \citep{2013A&A...558A..33A,2018AJ....156..123A},  Period04 \citep{2005CoAst.146...53L}
          }

\bibliography{example}{}
\bibliographystyle{aasjournal}

\clearpage


\begin{table*}

    \centering
    \begin{tabular}{lcccccc}
        \hline
       HJD-2,450,000 & Exposure & S/N(5850 \AA) &  W$_{\lambda}$ H$\alpha$ & W$_{\lambda}$ H$\beta$ & W$_{\lambda}$ He I (5876) & W$_{\lambda}$ He I (6678) \\ 
       \hline
         8392.8792 & 1x900 s& 35  & $-21.54 \pm 0.758$ & $-2.01 \pm 0.224$ & $-0.39 \pm 0.688$ & $-0.04 \pm 0.101$ \\
         
         8393.8875 & 1x900 s& 23  & $-21.32 \pm 1.213$ & $-1.99 \pm 0.201$  & $-0.40 \pm 0.699$ & $-0.01 \pm 0.117$\\  
         
         8410.7982 & 3x900 s& 53 & $-21.68 \pm 0.740$ & $-1.98 \pm 0.201$ & $-0.41 \pm 0.035$ & $-0.00 \pm 0.094$\\ 
         
         8440.8389 & 3x900 s& 78  & $-21.94 \pm 0.468$ & $-2.10 \pm 0.136$ & $-0.39 \pm 0.650$ & $-0.04 \pm 0.085$ \\ 
         
         8452.8008 & 3x900 s& 67 & $-22.01 \pm 0.477$ & $-2.21 \pm 0.134$ & $-0.43 \pm 0.691$ & $-0.08 \pm 0.079$\\ 
         
         8466.7714 & 3x900 s& 89  & $-22.22 \pm 0.587$ & $-2.24 \pm 0.113 $ & $-0.45 \pm 0.669$ & $-0.07 \pm 0.088$ \\ 
         
         8482.7012 & 3x900 s& 67  & $-22.21 \pm 0.557$ & $-2.32 \pm 0.119$ & $-0.46 \pm 0.709$ & $-0.09 \pm 0.084$ \\ 
         
         8494.6486 & 3x900 s& 46 & $-22.15 \pm 0.915$ & $-2.26 \pm 0.144$ & $-0.46 \pm 0.726$ & $-0.09 \pm 0.085$	\\ 
         
         8509.6697 & 3x900 s& 32  & $-22.93 \pm 1.046$ & $-2.16 \pm 0.154$ & $-0.39 \pm 0.692$ & $-0.05 \pm 0.092$ \\ 
         
         8543.5658 & 3x900 s& 44  & $-22.04 \pm 1.149$ & $-2.03 \pm 0.118$ & $-0.38 \pm 0.680$ & $-0.01 \pm 0.085$ \\ 
         
         8551.6850 & 3x900 s& 40  & $-22.32 \pm 0.601$ & $-2.06 \pm 0.143$ & $-0.38 \pm 0.689$ & $-0.02 \pm 0.092$ \\ 
         
         8558.6199 & 3x900 s& 61 & $-22.47 \pm 0.625$ & $-1.99 \pm 0.143$ & $-0.40 \pm 0.689$ & $-0.03 \pm 0.086$ \\ 
         
         8564.5627 & 3x900 s& 52  & $-22.61 \pm 0.650$ & $-2.04 \pm 0.149$ & $-0.39 \pm 0.703$ & $-0.02 \pm 0.093$	 \\ 
         
         8582.5988 & 3x900 s& 58  & $-22.77 \pm 0.618$ & $-2.04 \pm 0.097$ & $-0.42 \pm 0.702$ & $-0.03 \pm 0.084$ \\ 
         
         8597.5355 & 3x900 s& 32 & $-22.45 \pm 0.745$ & $-2.06 \pm 0.155$ & $-0.40 \pm 0.701$ & $-0.06 \pm 0.100$ \\ 
         
         8624.5316 & 3x900 s& 50 & $-23.09 \pm 0.692$ & $-2.22 \pm 0.164$ & $-0.46 \pm 0.720$ & $-0.05 \pm 0.100$ \\ 
         
        9197.7618 & 2x900 s& 66  & $-24.55 \pm 0.670$ & $-2.28 \pm 0.152$ & $-0.44 \pm 0.685$ & $+0.02 \pm0.048$\\
        
        9208.7543 & 2x900 s& 35 & $-24.15 \pm 0.703$ & $-2.30 \pm 0.159$ & $-0.45 \pm 0.707$ & $-0.08 \pm 0.097$\\
        
        9209.7026 & 2x900 s& 46  & $-24.17 \pm 0.664$ & $-2.29 \pm 0.154$ & $-0.44 \pm 0.697$ & $-0.08 \pm 0.090$\\
        
        9219.7222 & 2x900 s& 64  & $-24.99 \pm 0.695$ & $-2.17 \pm 0.154$ & $-0.43 \pm 0.646$ & $-0.08 \pm 0.085$\\
        
        9224.6465 & 2x900 s& 74  & $-24.16 \pm 0.721$ & $-2.27 \pm 0.143$ & $-0.45 \pm 0.699$ & $-0.07 \pm 0.083$\\
        
        9228.7730 & 2x900 s& 81  & $-24.40 \pm 0.780$ & $-2.29 \pm 0.168$ & $-0.45 \pm 0.696$ & $-0.08 \pm 0.096 $ \\
        
        9234.6194 & 2x900 s& 32  & $-24.00 \pm 0.670$ & $-2.28 \pm 0.162$ & $-0.42 \pm 0.682$ & $-0.08 \pm 0.080$ \\
        
        9240.6824 & 2x900 s& 42  & $-24.06 \pm 0.686$ & $-2.27 \pm 0.143$ & $-0.44 \pm 0.686$ & $-0.08 \pm 0.083$\\
        
\hline
    \end{tabular}
    \caption{Measurements}
    \label{measurements}
\end{table*}

\end{document}